# High-Performance Flexible Magnetic Tunnel Junctions for Smart Miniaturized Instruments

*Selma. Amara[1], Gallo. A. Torres Sevilla[1], Mayyada. Hawsawi[1], Yousof. Mashraei[1], Hanan .Mohammed[1], Melvin E. Cruz[1], Yurii.P.Ivanov[1,2,3], Samridh. Jaiswal[4,5], Gerhard. Jakob[4], Mathias. Kläui[4], Muhammad. Hussain[1], Jurgen. Kosel[1]\*.*

[1] Computer Electrical and Mathematical Science and Engineering.

King Abdullah University of Science and Technology, Thuwal 23955, Saudi Arabia.

[2] Erich Schmid Institute of Materials Science, Austrian Academy of Sciences, Jahnstrasse 12, A-8700, Leoben, Austria.

[3] School of Natural Sciences, Far Eastern Federal University, 690950, Vladivostok, Russia.

[4] Institut für Physik, Johannes Gutenberg Universität Mainz, 55128 Mainz, Germany.

[5] Singulus Technologies AG, 63796 Kahl am Main, Germany.

jurgen.kosel@kaust.edu.sa

MANUSCRIPT TEXT

1. Introduction

Flexible electronics has become an established field that has seen tremendous developments over the last years. [1–10] The unique capability to adjust the geometry of devices to curved surfaces or surfaces of changing shape provides vast advantages over conventional electronics on rigid substrates. Hence, flexible printed circuit boards have already become an industrial standard for medical implants or consumer electronics [11–14], where the main requirements are: large area, extreme thinness, and conformity to curved surfaces. The recent progress of organic [6,15,16] as well as inorganic [14,17,18] electronics, which are basically performed using printing or thin film technologies, has been very beneficial for the improvement of flexing electronic devices. In combination with other methods like screen or inkjet printing, novel concepts for flexible electronics have been developed for several applications including displays, [19] organic light-emitting diodes, [20] sensors, [21–25] radio frequency identification tags, [26–28] and organic solar cells. [29] Meanwhile, there are increasing activities on flexible magnetic field sensors and magnetoelectronics, [30,31] due to ubiquitous use of such kinds of devices. [18,32] By now, high-performance magnetic sensors based on the giant magnetoresistive (GMR) effect and the tunneling magnetoresistive (TMR) effect are mainly prepared using thin film fabrication technologies.[18,33-39] This method allows for fabricating extremely sensitive magnetic sensors, especially on Si substrates, and it would therefore be a desirable solution to flex the silicon substrate, where the magnetic elements are fabricated onto. Magnetic tunnel junctions (MTJs) are of particular interest, because they constitute the main component of diverse spintronic devices such as spin-transfer oscillators, magnetic sensors, hard-disk-drive read-heads or Magnetic Random Access Memories.[40]

MTJs are made of a magnetic multilayer thin film material. For sensor applications, the low power consumption [41], high sensitivity and small size make them an attractive option. MTJs with amorphous Al-oxide barriers exhibit TMR ratios below 100 %. This can be improved by using (001)-oriented MgO barriers, achieving TMR ratios over 150 % at room temperature. [42–44] These high values can be exploited to design very sensitive TMR sensors, [45–50] such as biomagnetic field sensors [51]. They show an important change in resistance versus the magnetic field than other sensors such as those utilizing the anisotropic magnetoresistance or giant magnetoresistance. [52–54] Compared to conventionally used Hall effect and AMR sensors, a TMR sensor can also have higher sensitivity, lower power consumption [41], better temperature stability and, in particular, compared to an AMR sensor, a wider linear range can be obtained. [54]

Flexible MTJ devices have been reported with alumina [37,39] and MgO tunnel barriers [55]. Mostly, they have been fabricated on organic flexible substrates. [55-58] Recently, it has been demonstrated that MTJs with an MgO barrier can be grown on top of surface-treated polyester-based organic substrates, [37] while some other works addressed the possibility of integrating MTJ elements into bendable membranes. [59] Organic MTJ devices could benefit from the integration with polymeric electronic devices for flexible, all-organic electronic systems, but the effectiveness of processes like roll-to-roll configuration for low-cost implementation [60,61], in terms of performance, operation voltage, etc. are still very inferior compared to state-of-the-art silicon-based transistor electronics. In addition, the processes involved still pose many challenges. In addition, patterning magnetic devices on non-standard substrates is difficult, due to detrimental effects of etching processes or heating of the substrate. Also, structuring devices with only a few hundred nanometers in lateral dimensions onto a non-planar, polymeric substrate, such as aromatic polyimides, has not been achieved so far. It was demonstrated that Kapton, a polyimide, is a very promising flexible substrate for magneto-electronic applications. [62]

In this work, we demonstrate high performance flexible MTJ elements fabricated using a silicon substrate flexing approach. To this end, MTJ stacks were fabricated on a Si substrate, which was thinned down from the backside to obtain MTJs on flexible Si substrate. This method is attractive, because it utilizes the fabrication process that has been optimized for MTJ devices without compromising their performance. It is not limited in terms of high temperature treatments or large thermal budgets, which are obstacles, when using flexible polymer substrates. As an example for an application on a miniaturized instrument, the flexible MTJ sensors where implemented on a cardiac catheter.

2. Results and Discussion

2.1. Flexible MTJ films

In order to prepare the flexible MgO-barrier MTJs, a polished silicon wafer was used with a nominal thickness of 500 µm and a 300 nm thick $SiO_2$ layer on one side of the substrate. The stacks were prepared at room temperature. Metallic multilayers were grown by DC-magnetron sputtering, and the MgO layer was grown by RF-sputtering using a Singulus Rotaris deposition tool compatible with a scale-up to industrial fabrication. **Figure 1** outlines the fabrication process, which is further detailed in the experimental methods section. At the end of this process, we obtain an ultrathin MTJ stack on a 3-5 µm thin silicon substrate.

After transforming the rigid into a flexible MTJ multilayer stack, the coercive field of the un-patterned sample was measured for different bending curvatures using a vibrating sample magnetometer (VSM). **Figure 2.a** shows the coercive field, saturation and remanent magnetization obtained for several samples of the same wafer and for different bending diameter down to 500 µm. Note, a curvature of 0 corresponds to the measurements taken on a rigid sample. To this end, the samples were placed on a VSM holder (**Figure 2.b**), and the bending diameter was evaluated from scanning electron microscopy (SEM) images as shown in **Figure 2.c**. For all diameters, a similar coercivity was found with an average value of 22

Oe, minimum and maximum values of 17 Oe and 28 Oe, respectively, which confirms that the magnetic properties were not affected or the structure damaged during the transformation process of the rigid MTJ into a flexible one. The saturation and the remanence magnetizations show the same behavior, i.e. independently of the bending diameters, they are stable with an average value of $9\times10^{-6}$ and $2\times10^{-6}$ emu, respectively. The structure of the MTJ multilayers was analyzed by transmission electron microscopy after the flexing process. As can be seen in **Figure 3**, the flexible sample shows no structural degradation compared to its rigid, bulk counterpart, and the well-defined nanoscale multilayer structure is entirely maintained.

By attaching a flexible sample to a VSM holder with the MTJ stack facing inward or outward, compressive and tensile strains, respectively, can be applied to the MTJ film and the magnetic properties can be measured with the VSM tool. The corresponding magnetization curves are shown in **Figure 4.a** for MTJ films that are flat or bent in a compressive and tensile manner with a diameter of 2 mm. For the magnetic field oriented along the easy axis, a tensile strain along the easy axis gives rise to a harder magnetic behavior shifting the saturation field from around 2000 Oe to 2500 Oe. In contrast, under a compressive strain, the material becomes magnetically softer. Therefore, this indicates that the total magnetostriction [63] of the TMR stack is negative. Those results provide an alternative way to mechanically tune the magnetic properties, which should be considered, when developing flexible magnetic devices.

In order to evaluate the reliability of the flexible MTJs, their magnetic properties were studied under periodic strain cycles. This was accomplished by attaching a sample to an elastomeric support, which was repetitively stressed by bringing it from a flat state into a tensile state and visa-versa. After 10, 100 and 1000 of such cycles, the magnetization curves were measured and no relevant differences, due to fatigue, were found, as shown in **Figure 4.b**.

2.2. Magnetoresistance of flexible patterned MTJs

The resistance of an MTJ depends on the relative orientation of the magnetization directions of the two magnetic layers to each other, due to spin-dependent tunneling involved in the transport between the majority and minority spin states. [64]

It reaches a maximum value when the two layers are antiparallel and a minimum when they are parallel. This resistance change is typically quantified by the TMR ratio defined as:

$$TMR(\%) = \frac{[R_{Max} - R_{Min}]}{R_{Min}}$$

Where $R_{Max}$ and $R_{Min}$ correspond to the resistance for antiparallel and parallel magnetization configurations between the two magnetic layers, respectively. We used the TMR ratio to characterize the performance of patterned MTJs before and after thinning the Si substrate, i.e. before and after transforming the rigid MTJs into flexible ones.

For static testing, the MTJ elements were attached to cylindrical rods of specific diameters. **Figure 5.b** shows the TMR ratio together with the values of $R_{Max}$ and $R_{Min}$ versus the bending diameter. All values remain constant for bending diameters from 30 to 3 mm with roughly $R_{Max} = 22\ \Omega$, $R_{Min} = 9\ \Omega$ and a TMR ratio = 145 %.

All measurements show similar behaviors for different bending conditions, which suggests that the stress applied to the MgO-barrier MTJs is not large enough to change the electron band structure. This is consistent with the TMR behavior, which shows almost no change for different bending conditions.

As before, cyclic loading experiments were performed to test the reliability of the bendable MTJ elements. TMR ratios were taken before applying (zero cycles) and after the $10^{th}$, $100^{th}$ and $1,000^{th}$ loading cycle. As displayed in **Figure 5.d**, the flexible MTJ elements do not show any signs of fatigue.

3. Flexible MTJs Sensors for Smart Miniaturized Instruments

Flexible MTJ sensors are ideal for applications on smart miniaturized instruments, where minimal changes to the dimensions and weight are essential and need to be combined with a high sensing performance. An example is catheter tracking during cardiac catheterization procedures, which are performed either for the treatment or diagnosis of the cardiovascular system. During such procedures, x-ray imaging and contrast agents are commonly required to visualize and track the catheter inside the heart, as shown in **Figure 6.a** and **Figure 6.b**. Although cardiac catheterization is a minimally invasive surgery, orientation tracking of the catheter can be cumbersome and, in many cases, require several attempts to assure that the catheter is in the correct position and orientation inside the heart, leading to excessive use of contrast agents and x-ray doses. Negative implications are associated with both, the x-ray exposure and the contrast agent [65, 66]. Contrast induced nephropathy (CIN) is a failure in the renal functions caused by the intravenous contrast agents, which can lead to acute rental dys-function[67]. Several solutions have been proposed to minimize these side effects; however, they either fail to fulfill the miniaturization requirement [68] or offer bulky systems that require special installations in the laboratory [69,70]. Hence, the fabricated flexible MTJ sensors can be the starting point towards the development of a highly miniaturized catheter system for orientation tracking with minimal side effects.

A flexible MTJ sensor was attached to the tip of a cardiac catheter with a diameter of 2 mm using a LOCTITE 4011 super glue, as shown in **Figure 6.c**. The resistance of the MTJ sensor was measured using 4 points probes and a Keithly 2400-C in 4-wire resistance mode. A magnetic field was applied with Helmholtz coils from -50 Oe to 50 Oe with 0.1 Oe steps. A rotary table made of a laser-cut PMMA was designed to obtain the MTJ resistance as a function of the angle of the applied magnetic field. Thereby, 0° and 90° correspond to the direction of the easy axis and hard axis of the MTJ sensor, respectively. As shown in **Figure 6.d**, the flexible MTJ sensor maintained its performance after attaching it to the catheter and shows a clear response to different angles of the magnetic field, which could be further exploited as part of an orientation monitoring system.

The fabricated MTJ sensors are particularly suitable for catheter tracking, due to their small size, low power consumption and high sensitivity, which are 150 µm$^2$, 0.15 µW and 4.93 %/Oe, respectively. The sensor only adds 8 µg to the weight of the catheter and 15 µm to its diameter. The capability to bend it up to 500 µm, is sufficient even for the smallest catheters in use with a diameter of 1 mm (3F). It is also worth noting that the sensors are sensitive to magnetic fields amplitudes of less than 1 Oe, indicating that the Earth's magnetic field could be utilized for orientation tracking.

4. Conclusion

We demonstrate a method to fabricate flexible Magnetic Tunnel Junctions on Si substrate with high performance, flexibility and mechanical endurance. The flexible MTJs were fabricated from standard multilayer magnetic stacks deposited on a Si wafer. By backside etching of the wafer the thickness of the substrate was reduced from 500 µm to 5 µm, which is three times thinner than the thinnest recently reported flexible MTJs on silicon substrate [71]. The ultra-thin substrate enables these flexible MTJ films to be bent with curvatures down to 500 µm, making them suitable for conformal fitting onto the surface of miniaturized devices, like surgical instruments. We found that using this process maintains the magnetic properties and magnetoresistive characteristics of the flexible MTJs, without degrading the performance in terms of TMR ratio compared to their rigid bulk counterpart. A high reliability was also found for the MTJ elements, which endured over 1,000 bending cycles without fatigue.

This method enables fabricating devices without comprising performance, integration density and cost/yield advantage of silicon microfabrication processes. It has the advantage of allowing for the fabrication of entire devices prior to thinning and keeps standard processes without introducing any constraint in design and commonly known deposition, etching and lithography methods. Also, thermal budget constraints are kept intact when applying industry standard processes. Hence, the flexible MTJs presented in this paper are a major step toward the integration of state-of-the-art flexible magnetoelectronics.

Implementing the flexible MTJ sensors at the tip of a cardiac catheter with 2 mm in diameter did not compromise their performance. The capability to detect magnetic fields with a high sensitivity of 4.93 %/Oe with a device of 150 µm$^2$ in size and 8 µg in weight shows the potential of this method for creating a new generation of miniaturized instruments with embedded electronics. An added thickness of 5 µm to the diameter of the catheter can be considered negligible. The low power consumption of 0.15 µW is another attractive feature that avoids unwanted heating of the device and would allow wireless operation.

4. Methods Section

Device Fabrication:

The studied MTJs were deposited onto thermally oxidized Si (001) wafers using an ultrahigh vacuum ($p_{base} < 8 \cdot 10^{-7}$ Pa) magnetron sputtering system (Singulus Rotaris) able to deposit homogeneously on 8" wafers. The stacking structure of the films was:

Si/SiO$_2$/Ta(5)/CuN(25)/Ta(5)/CuN(25)/Ta(3)/Pt$_{38}$Mn$_{62}$(20)/Co$_{70}$Fe$_{30}$(2.2)/Ru(0.85)/ Co$_{40}$Fe$_{40}$B$_{20}$(2.4)/MgO(1.6)/Co$_{40}$Fe$_{40}$B$_{20}$(2.4)/Ta(10)/CuN(30)/Ru(7)Ta10) (in nm).

The magnetoresistance properties were measured at RT using four probes with an applied magnetic field on patterned MTJs. They were of elliptical pillars with 1 µm * 1.2 µm patterned by photolithography and Ar ion milling showing 120 %–150 % TMR ratio and a resistance area product of 10 - 12 Ω·µm$^2$.

Flexing process:

The flexible MTJs were made using a standard MTJ fabrication process on a silicon substrate followed by thinning of the substrate, as shown in **Figure 1**. The Si (500 µm) / SiO$_2$ (300 nm) substrate was thinned by a soft-etch-back method [72,73], transforming the microscale MTJs into flexible ones. First, the top surface of the devices was covered with a protective polymer by spinning photoresist of 7 µm thickness to prevent degradation of the devices in the subsequent etching steps. Then, the devices were processed by deep-reactive-ion-etching from the backside, to reduce the thickness of the silicon substrate from 500 µm to 5 µm. To facilitate a reliable etching process and prevent damage of the devices, due to over etching, the complete process was divided into four etching stages. The first step reduced the thickness of the wafer from 500 µm to 200 µm. Next, each of the remaining etching steps reduced the thickness by 50 µm. Once the thickness of 50 µm was reached, the etching steps were modified to reduce the thickness by 5 µm until a final thickness of 5 µm was obtained. At this point, the MTJ stacks were flexible with down to 500 µm naturally bending diameter. It is to be noted that two different profiler measurements were performed after each etching step to ensure etching uniformity and prevent degradation of the MTJ sensors. The first profiler measurement was based on a mechanical profilometer that uses a stylus to measure the step size between the back surface of the sample and the carrier wafer. The second measurement was based on an optical profilometer, which uses reflected light to measure the step size between two different surfaces. The difference between the mechanical and the optical profilometer measurements was found to be ±3 µm, which confirms the uniformity during the etching process.

Magnetoresistance measurements:

The magnetoresistance properties of the junctions were measured at room temperature in air by a conventional DC four-probe method and current driven Helmholtz coils controlled with LabView. A current of 10 mA was applied using a Keithley 2400-C in 4-wire resistance mode and ten consecutive voltage measurements were averaged at each field step. The iron-core Helmholtz coils were driven by a Kepco Bop72-6m Bipolar DC Power Supply / Amplifier (072v 6a 400W) working as amplifier to another Keithley 2400-C in current source mode, while the field was monitored using a F.W. Bell Gaussmeter model 6010. The field was swept with 0.1 Oe steps, and its direction was the same as that of the easy axis of the free layer.

The magnetization curves of the as-deposited samples were measured by a vibrating sample magnetometer (Lakeshore 7404).

For static bending experiments, the flexible samples were fixed with a double-sided adhesive tape onto the faces of PMMA cylinders with different diameters ranging from 3mm to 30 mm.

TEM studies:

Electron microscopy studies were performed on a Cs corrected (scanning) transmission electron microscope Titan 60-300 (FEI, Netherlands), operated at 200 kV. Cross-sections of MTJ multilayer stacks before and after the thinning process were prepared using a focused ion beam. The EELS experiments were performed with a post-column high resolution Gatan energy filtering spectrometer operated at a dispersion of 1 eV. The optical conditions of the microscope for EELS imaging and spectroscopy were set to obtain a probe-size of 0.2 nm with a convergence semi-angle of 10 mrad and a collection semi-angle of 12 mrad.

Dynamic bending studies:

The bendability of the TMR-sensing elements was tested using an automated computer-controlled bending stage in conjunction with a four-point resistance measurement. The samples were attached to a 1 mm thick Kapton tape and repetitively stressed by bringing them from a flat state into a curved state with 3 cm in diameter (tensile strain) and visa-versa.

ACKNOWLEDGMENT

Research reported in this publication was supported by the King Abdullah University of Science and Technology (KAUST), the European Research Council through a Proof-of-Concept grant (MultiRev ERC-2014-PoC (665672)), the European Community under the Marie-Curie Seventh Framework program – ITN "WALL" (Grant No. 608031), as well as the German research foundation (DFG) and the State Research Center of Innovative and Emerging Materials at Johannes Gutenberg-University Mainz (CINEMA).

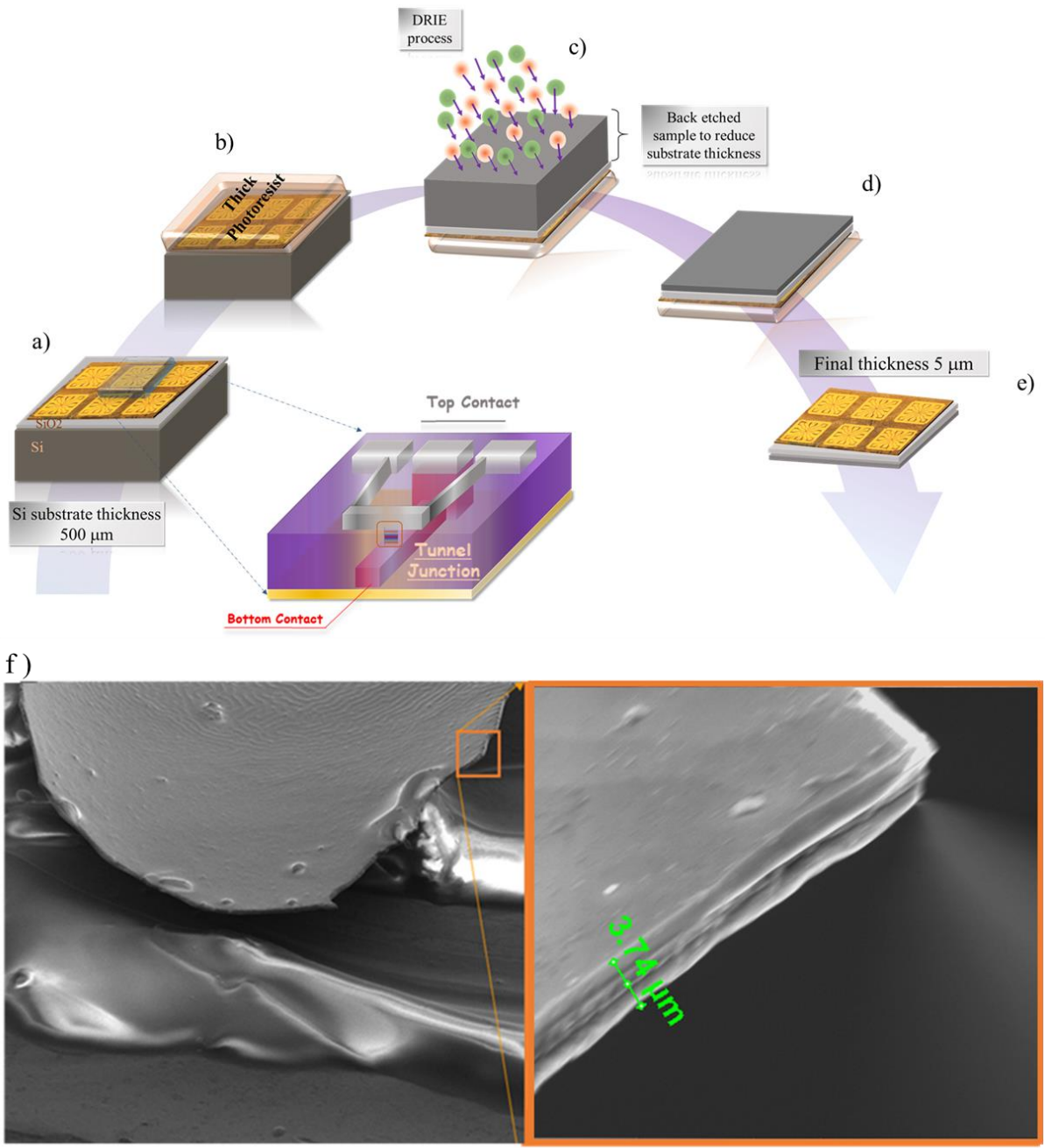

**Figure 1.** Fabrication process flow: (a) MTJ devices fabricated on SiO$_2$ (300 nm) / Si (500 μm) substrate; (b) Photoresist coating for chip-protection during subsequent back etching process; (c) Etching of the substrate's backside using deep reactive ion etching (DRIE); (d) MTJ sensor devices on 3-5 μm thick flexible silicon substrate; and (e) Photoresist removal.
f) SEM image of MTJ stack on flexible Si substrate showing the final thickness after the flexing process.

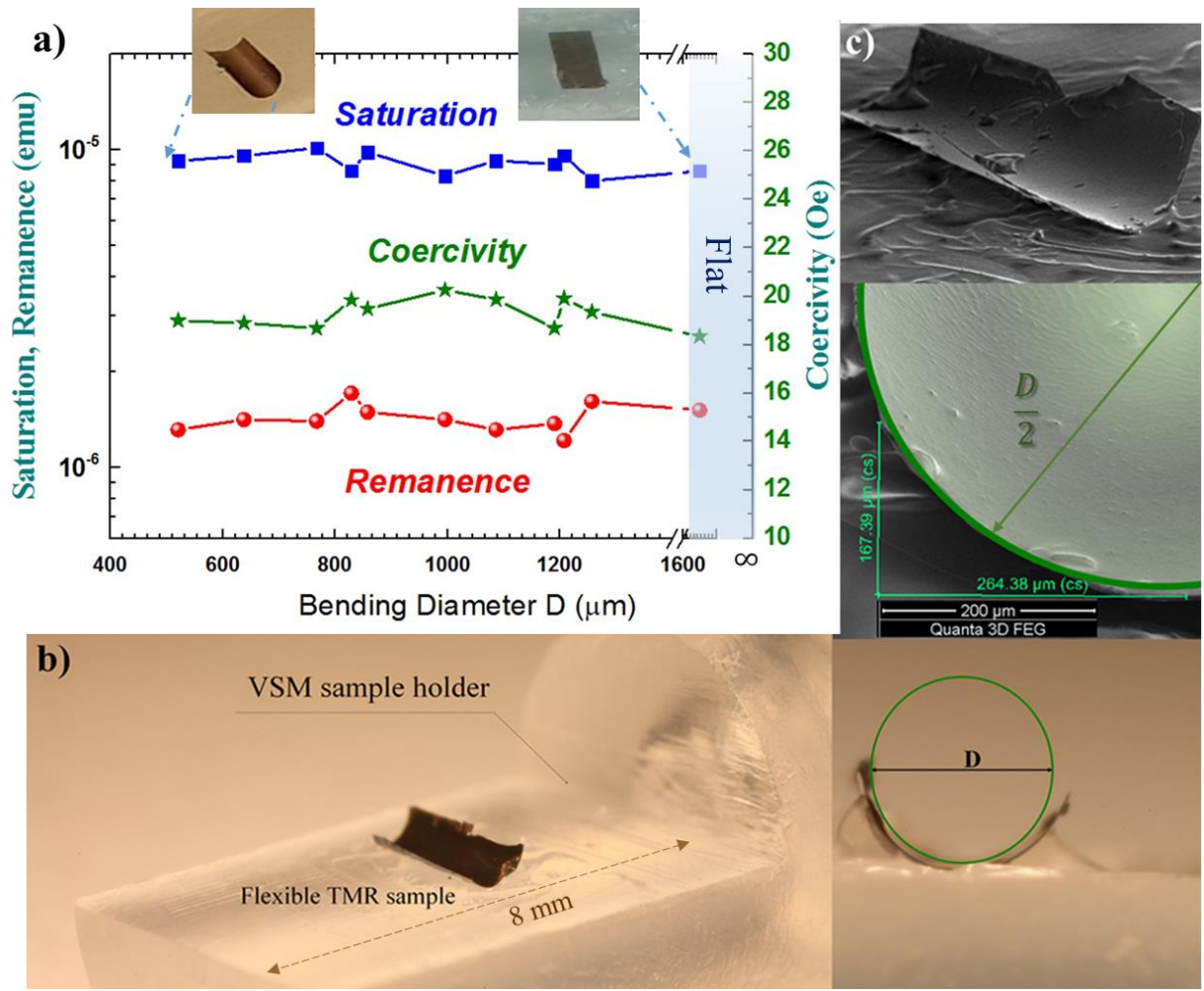

**Figure 2.** (a) Coercive field, saturation and remanent magnetizations of the MTJ stack on flexible Si substrate measured with a vibrating sample magnetometer versus the bending curvature (1/diameter). (b) Optical microscope image of a curved MTJ stack on flexible Si substrate on top of the VSM sample holder. (c) SEM image of MTJ stack on flexible Si substrate and measurement method for the bending diameter (D).

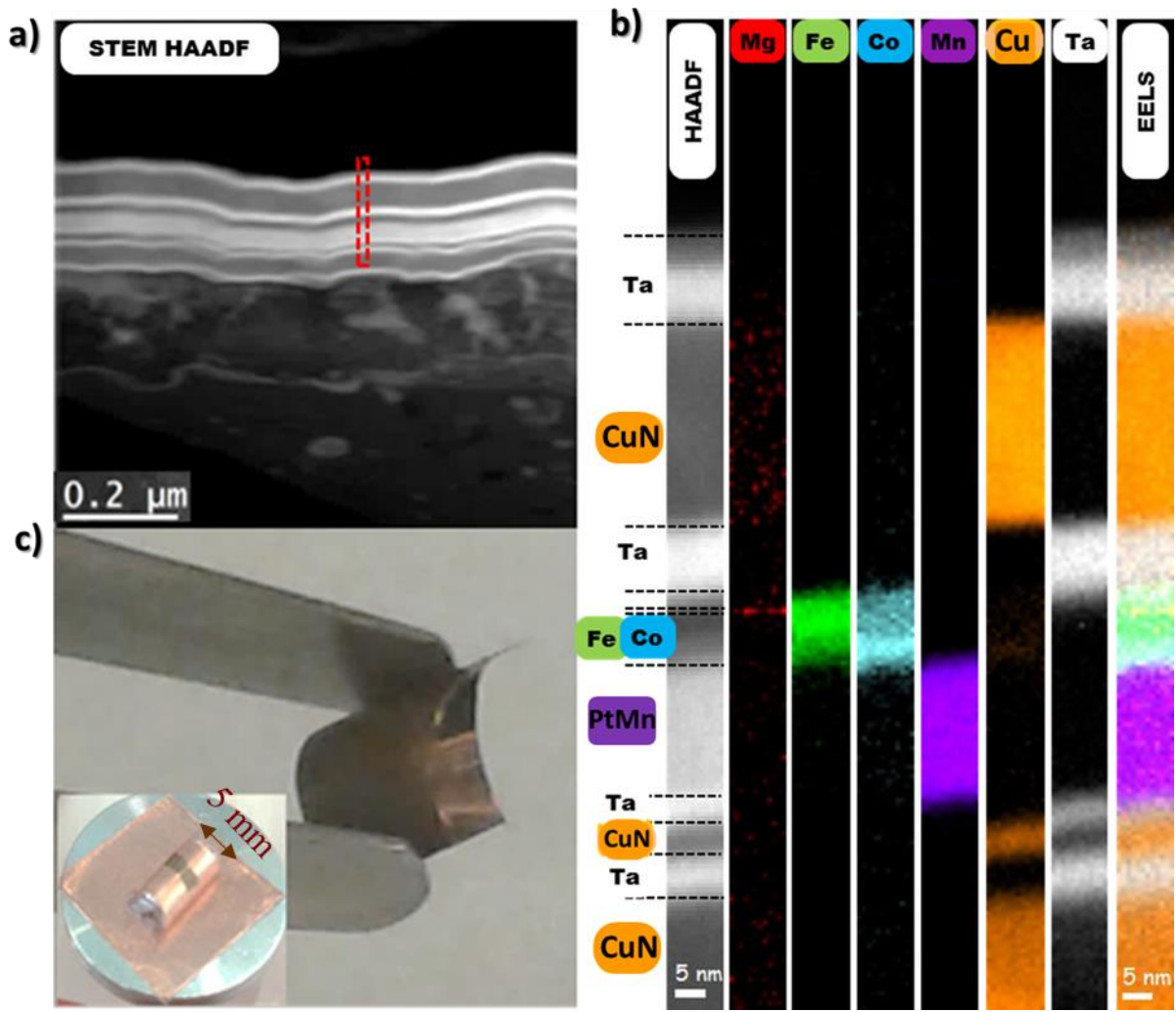

**Figure 3.** (a) Scanning Transmission Electron Microscopy image after reducing the thickness of the Si substrate from 500 μm to 5 μm. (b) The different layers of the magnetic stack on the flexible sample are intact after the thinning process and no damages are found using the EELS elemental analysis. (c) Example of flexible MTJ film.

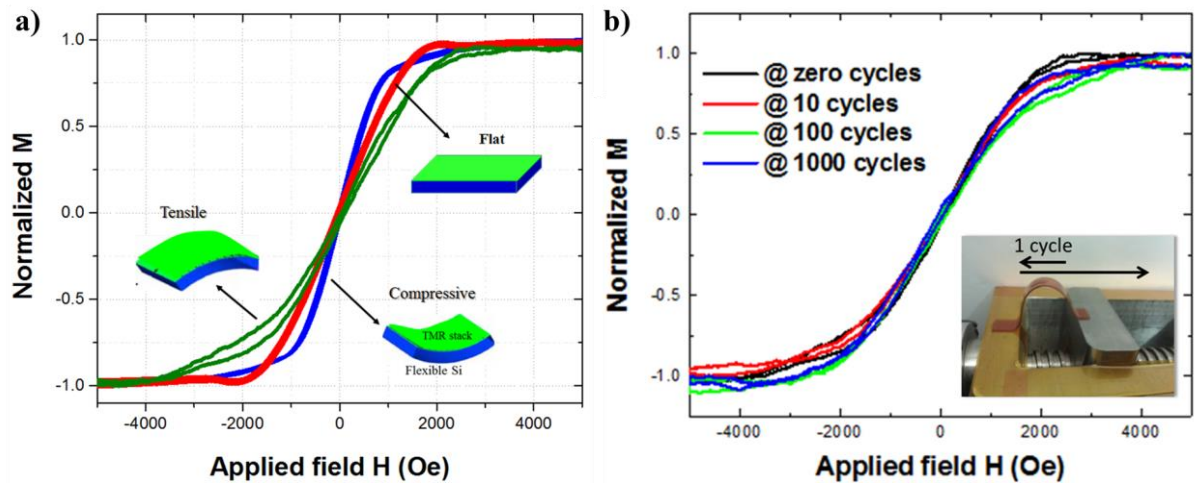

**Figure 4.** (a) Magnetization loops of an MTJ stack after the flexing process, when the sample is on a flat or curved surface with the stack either facing the surface (compressive stress) or facing away from it (tensile stress). (b) Magnetization curves of a flexible MTJ before mechanical stressing and after 10, 100 and 1000 cycles of bending with tensile stress.

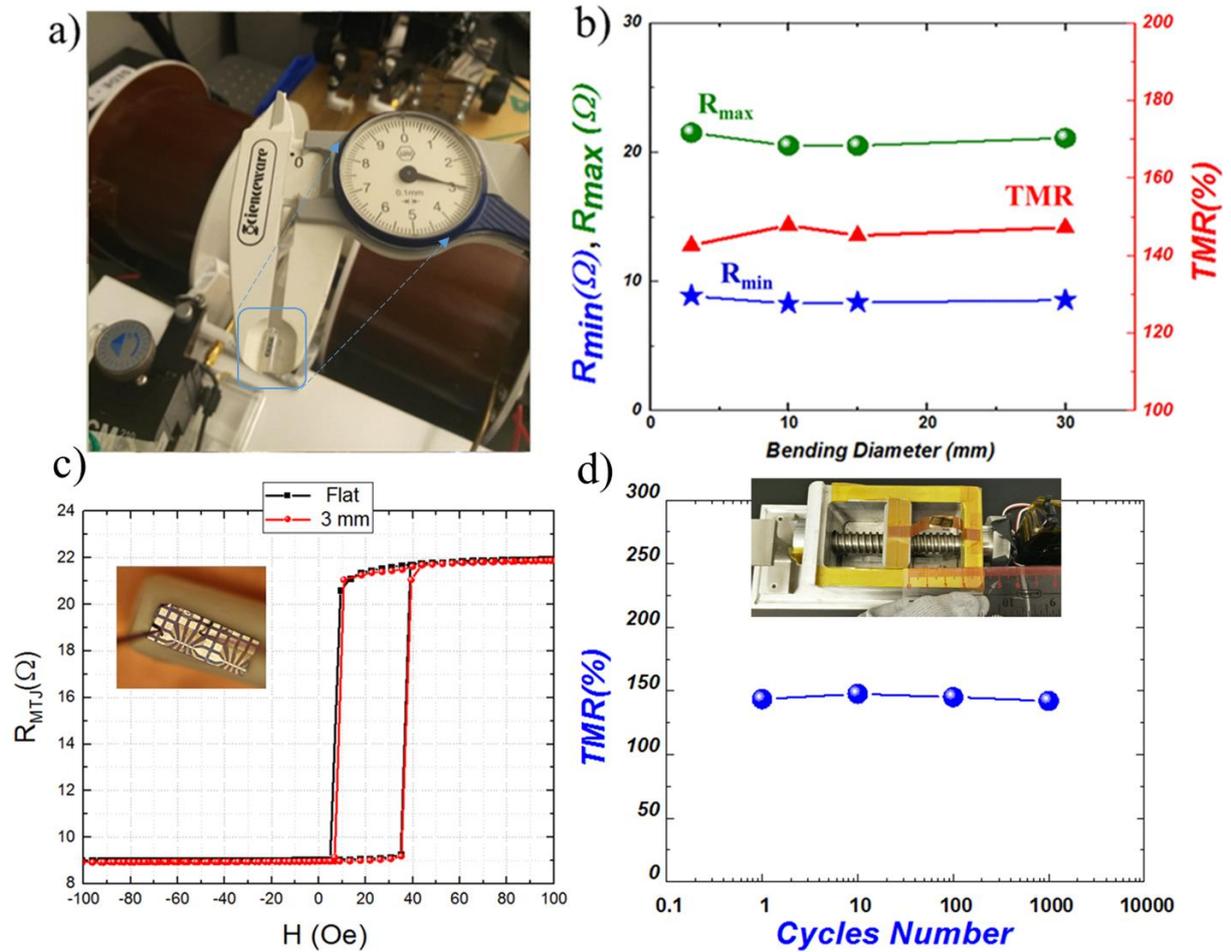

**Figure 5.** (a) Image of a flexible MTJ sensor. A die with four MTJs is fixed on top of a cylinder with 3 mm in diameter (tensile strain). (b) Resistance in the parallel state ($R_{min}$) and antiparallel state ($R_{max}$), as well as TMR ratios for a positive bending diameter from 30 mm to 3 mm. The junction size is $1\times1.2$ μm$^2$ and the RA product is about 10 Ωμm$^2$. (c) Example of TMR loops of MTJ in flat state and when bent with a diameter of 3 mm. (d) TMR ratio versus cycles of tensile strain. The inset shows the experimental setup, which was used to cyclically apply the strain.

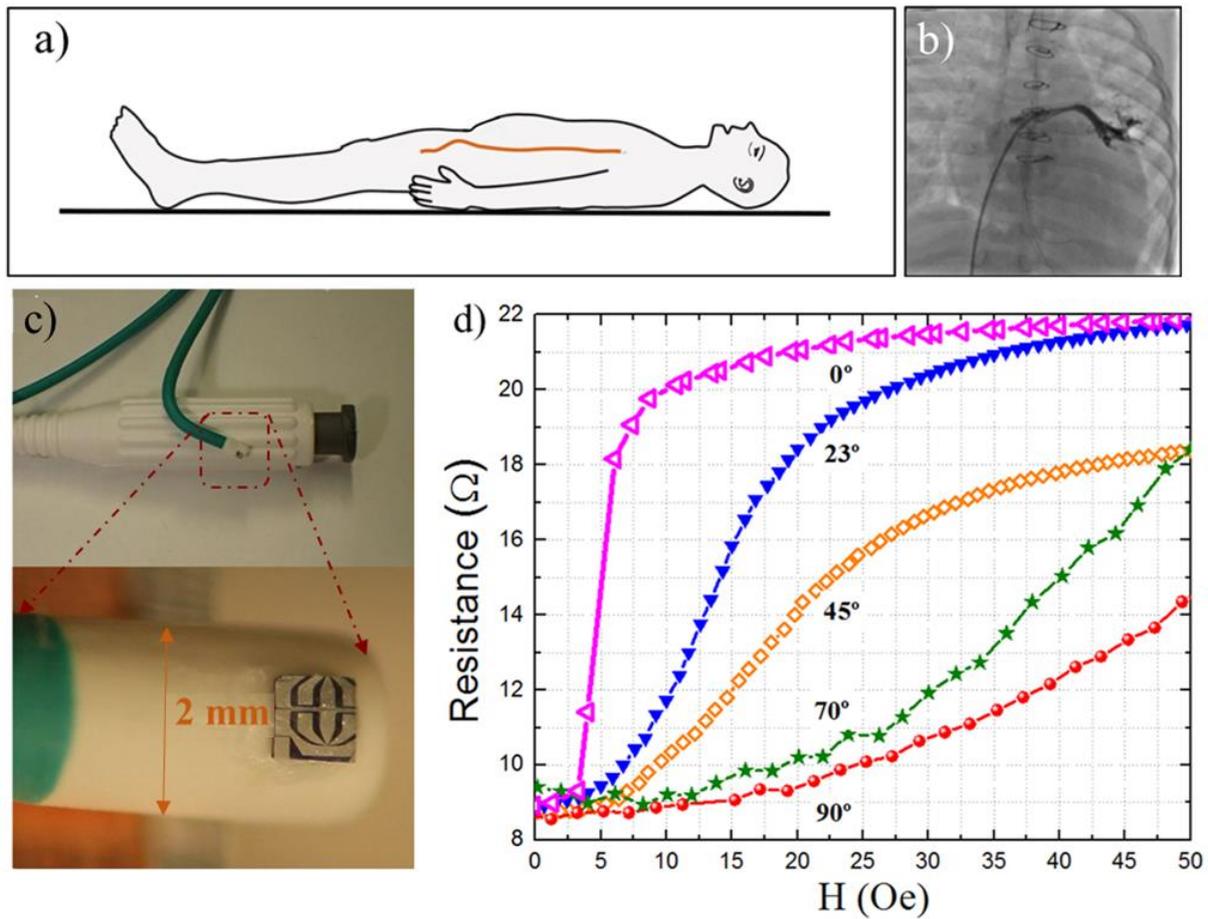

**Figure 6.** (a) In cardiac catheterization, a cardiac catheter is inserted into the heart of a patient through his groin. (b) An angiographic image of a patient showing the catheter and the contrast agent. [74]. (c) The attachment of flexible MTJ sensors onto the tip of a diagnostic cardiac catheter (6F). d) Resistance response of a flexible MTJ sensor attached to the catheter tip for different angles of the applied magnetic field.